\documentstyle[aps,prb,epsf,twocolumn]{revtex}

\begin{document}
\draft
\title{Phase-dependent magnetoconductance fluctuations in a chaotic Josephson
junction}

\author{P. W. Brouwer and C. W. J. Beenakker}
\address{Instituut-Lorentz, University of Leiden, P.O. Box 9506, 2300 RA
Leiden, The Netherlands\medskip \\ \parbox{14cm}{\rm
Motivated by recent experiments by Den Hartog et al., we present a
random-matrix theory for the  magnetoconductance fluctuations of a chaotic
quantum dot which is coupled by point contacts to two superconductors and one
or two normal metals. There are aperiodic conductance fluctuations as a
function of the magnetic field through the quantum dot and $2\pi$-periodic
fluctuations as a function of the phase difference $\phi$ of the
superconductors. If the coupling to the superconductors is weak compared to the
coupling to the normal metals, the $\phi$-dependence of the conductance is
harmonic, as observed in the experiment. In the opposite regime, the
conductance becomes a random $2\pi$-periodic function of $\phi$, in agreement
with the theory of Altshuler and Spivak. The theoretical method employs an
extension of the circular ensemble which can describe the magnetic field
dependence of the scattering matrix.\smallskip}}
\maketitle

\narrowtext
The conductance of a mesoscopic metal shows small fluctuations of universal
size $e^2/h$ as a function of magnetic field.\cite{AltshulerLeeStone} These
universal conductance fluctuations are sample-specific, which is why a plot of
conductance $G$ versus magnetic field $B$ is called a ``magnetofingerprint''.
The magnetoconductance is sample-specific because it depends sensitively on
scattering phase shifts, and hence on the precise configuration of scatterers.
Any agency which modifies phase shifts will modify the magnetoconductance.
Altshuler and Spivak\cite{AltshulerSpivak} first proposed to use a Josephson
junction for this purpose. If the metal is connected to two superconductors
with a phase difference $\phi$ of the order parameter, the conductance
$G(B,\phi)$ contains two types of sample-specific fluctuations: aperiodic
fluctuations as a function of $B$ and $2\pi$-periodic fluctuations as a
function of $\phi$. The magnetic field should be sufficiently large to break
time-reversal symmetry, otherwise the sample-specific fluctuations will be
obscured by a much stronger $B$- and $\phi$-dependence of the ensemble-averaged
conductance.\cite{BeenakkerLesHouches,Vegvar}

In a recent Letter, Den Hartog et al.\cite{Hartog}\ reported the experimental
observation of phase-dependent magnetoconductance fluctuations in a T-shaped
two-dimensional electron gas. The horizontal arm of the T is connected to two
superconductors, the vertical arm to a normal metal reservoir. The observed
magnitude of the fluctuations was much smaller than $e^2/h$, presumably because
the motion in the T-junction was nearly ballistic. Larger fluctuations are
expected if the arms of the T are closed, leaving only a small opening (a point
contact) for electrons to enter or leave the junction. Motion in the junction
can be ballistic or diffusive, as long as it is chaotic the statistics of the
conductance fluctuations will only depend on the number of modes in the point
contacts and not on the microscopic details of the junction.

In this paper we present a theory for phase-dependent magnetoconductance
fluctuations in a chaotic Josephson junction. We distinguish two regimes,
depending on the relative magnitude of the number of modes $M$ and $N$ in the
point contacts to the superconductors and normal metals respectively. For $M
\gg N$ the $\phi$-dependence of the conductance is strongly anharmonic. This is
the regime studied by Altshuler and Spivak.\cite{AltshulerSpivak} For $M
\lesssim N$ the oscillations are nearly sinusoidal, as observed by Den Hartog
et al.\cite{Hartog} The difference between the two regimes can be understood
qualitatively in terms of interfering Feynman paths. In the regime $M \lesssim
N$ only paths with a single Andreev reflection contribute to the conductance.
Each such path depends on $\phi$ with a phase factor $e^{\pm i \phi/2}$.
Interference of these paths yields a sinusoidal $\phi$-dependence of the
conductance. In the opposite regime $M \gg N$, quasiparticles undergo many
Andreev reflections before leaving the junction. Hence higher harmonics appear,
and the conductance becomes a random $2\pi$-periodic function of $\phi$.

\begin{figure}[h]
\vspace{-0.22cm}

\hspace{0.05\hsize}
\epsfxsize=0.8\hsize
\epsffile{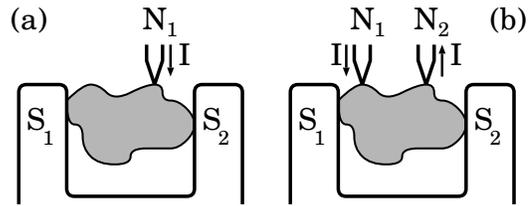}
\vspace{0.3cm}

\caption{\label{fig:1} Josephson junction in a three-terminal (a) and
four-terminal (b) configuration.}
\end{figure}

The system under consideration is shown schematically in Fig.\ \ref{fig:1}. It
consists of a chaotic cavity in a time-reversal-symmetry breaking magnetic
field $B$, which is coupled to two superconductors and to one or two normal
metals by ballistic point contacts. The superconductors (S$_1$ and S$_2$) have
the same voltage (defined as zero) and a phase difference $\phi$. The
conductance of this Josephson junction is measured in a three- or four-terminal
configuration. In the three-terminal configuration (Fig.\ \ref{fig:1}a), a
current $I$ flows from a normal metal N$_1$ into the superconductors. The
conductance $G = I/V_1$ is the ratio of $I$ and the voltage difference $V_1$
between N$_1$ and S$_1$, S$_2$. This corresponds to the experiment of Den
Hartog et al.\cite{Hartog} In the four-terminal configuration (Fig.\
\ref{fig:1}b), a current $I$ flows from a normal metal N$_1$ into another metal
N$_2$. The conductance $G=I/(V_1-V_2)$ now contains the voltage difference
between N$_1$ and N$_2$. This is the configuration studied by Altshuler and
Spivak.\cite{AltshulerSpivak}

Following Ref.\ \onlinecite{Hartog} we split the conductance $G(B,\phi) =
G_0(B) + G_{\phi}(B,\phi)$ into a $\phi$-independent background
\begin{eqnarray}
   G_0(B) &=& \int_{0}^{2 \pi}{ d\phi \over 2 \pi} \, G(B,\phi),
\end{eqnarray}
plus $2\pi$-periodic fluctuations $G_{\phi}$. In the absence of time-reversal
symmetry, the ensemble average $\langle G(B,\phi) \rangle \equiv \langle G
\rangle$ is independent of $B$ and $\phi$. Hence $\langle G_0(B) \rangle =
\langle G \rangle$ and $\langle G_{\phi}(B,\phi) \rangle = 0$. The correlator
of $G$ is
\begin{eqnarray}
  C(\delta B, \delta \phi) &=& \langle G(B,\phi) G(B + \delta B,\phi + \delta
\phi)\rangle - \langle G \rangle^2.
\end{eqnarray}
Fluctuations of the background conductance are described by the correlator of
$G_0$,
\begin{eqnarray}
  C_0(\delta B) &=&  \langle G_0(B) G_0(B+\delta B) \rangle - \langle G
\rangle^2 \nonumber \\ &=& \int_{0}^{2 \pi} {d \delta \phi \over 2 \pi} \,
C(\delta B, \delta \phi).
\end{eqnarray}
(In the second equality we used that $\langle G_{\phi} G_0 \rangle = 0$.) The
difference $C_{\phi} = C - C_0$ is the correlator of $G_{\phi}$,
\begin{eqnarray}
  C_{\phi}(\delta B, \delta \phi) &=& \langle G_{\phi}(B,\phi) G_{\phi}(B +
\delta B,\phi + \delta \phi)\rangle.
\end{eqnarray}
We compute these correlators for the three- and four-terminal configurations,
beginning with the former.

In the three-terminal configuration, the cavity is connected to three point
contacts (Fig.\ \ref{fig:1}a). The contact to the normal metal has $N$
propagating modes at the Fermi energy, the contacts to the superconductors have
$M/2$ modes each. The $(N+M) \times (N+M)$ scattering matrix $S$ of the cavity
is decomposed into $M \times M$ ($N \times N$) reflection matrices $r$ ($r'$)
and $N \times M$ ($M \times N$) transmission matrices $t$ ($t'$),
\begin{equation}
  S = \left( \begin{array}{cc} r & t' \\ t & r' \end{array} \right).
\end{equation}
The conductance at zero temperature is determined by the matrix $s_{he}$ of
scattering amplitudes from electron to hole,\cite{TakaneEbisawa,Beenakker92}
\begin{mathletters} \label{eq:NSconductance}
\begin{eqnarray} \label{eq:NSconductance1}
  G  &=& 2\, \mbox{tr}\, s_{he}^{\vphantom{dagger}} s_{he}^{\dagger}, \\
  s_{he}{\vphantom{dagger}} &=& -i\, t^{*} \left(1 + e^{i \Phi} r e^{-i \Phi}
r^{*}\right)^{-1}\! e^{i \Phi} t'. \label{eq:NSconductance2}
\end{eqnarray}
\end{mathletters}%
The diagonal matrix $\Phi$ has diagonal elements $\Phi_{nn} = \phi/2$ if $1 \le
n \le M/2$ and $-\phi/2$ if $1+M/2 \le n \le M$. We measure $G$ in units of
$2e^2/h$.

\begin{figure}[h]
\hspace{0.05\hsize}
\epsfxsize=0.8\hsize
\epsffile{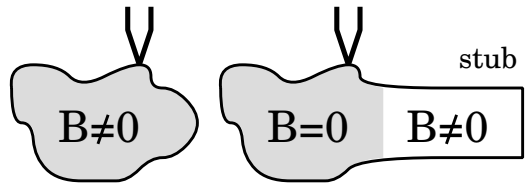}

\vspace{0.3cm}

\caption{\label{fig:2} Schematic picture how the magnetic field is included in
the scattering-matrix ensemble. A chaotic cavity with a spatially homogeneous
magnetic field (left diagram) is statistically equivalent to a chaotic cavity
in zero magnetic field (right diagram), which is coupled to a closed lead (a
stub) having a non-symmetric reflection matrix.}
\end{figure}

For chaotic scattering without time-reversal symmetry, the matrix $S$ is
uniformly distributed in the unitary group.\cite{BlumelSmilansky} This is the
circular unitary ensemble (CUE) of random-matrix theory.\cite{Mehta} The CUE
does not specify how $S$ at different values of $B$ is correlated. The
technical innovation used in this work is an extension of the CUE, which
includes the parametric dependence of the scattering matrix on the magnetic
field. The method (described in detail elsewhere\cite{future}) consists in
replacing the magnetic field by a time-reversal-symmetry breaking stub (see
Fig.\ \ref{fig:2}). This idea is similar in spirit to B\"uttiker's method of
modeling inelastic scattering by a phase-breaking lead.\cite{Buettiker} The
stub contains $N_{\rm stub}$ modes. The end of the stub is closed, so that it
conserves the number of particles without breaking phase coherence.
(B\"uttiker's lead, in contrast, is attached to a reservoir, which conserves
the number of particles by matching currents, not amplitudes, and therefore
breaks phase coherence.) We choose our scattering basis such that the $N_{\rm
stub} \times N_{\rm stub}$ reflection matrix $r_{\rm stub}(B)$ of the stub
equals the unit matrix at $B=0$. For non-zero magnetic fields we take
\begin{equation} \label{eq:VB}
  r_{\rm stub}(B) =  e^{B A}, \ \ a^2 \equiv \sum_{n < m} A_{nm}^2,
\end{equation}
where the matrix $A$ is real and antisymmetric: $A_{nm}^{\vphantom{*}} =
A_{nm}^{*} = - A_{mn}^{\vphantom{*}}$. Particle-number is conserved by the stub
because $r_{\rm stub}$ is unitary, but time-reversal symmetry is broken,
because $r_{\rm stub}$ is not symmetric if $B \neq 0$. In order to model a
spatially homogeneous magnetic field, it is essential that $N_{\rm stub} \gg N
+ M$. The value of $N_{\rm stub}$ and the precise choice of $A$ are irrelevant,
all results depending only on the single parameter $a$.

The magnetic-field dependent scattering matrix $S(B)$ in this model takes the
form
\begin{equation} \label{eq:CUEB}
  S(B) = U_{11} + U_{12} \left[1 - r_{\rm stub}(B) U_{22} \right]^{-1} r_{\rm
stub}(B) U_{21}.
\end{equation}
The matrices $U_{ij}$ are the four blocks of a matrix $U$ representing the
scattering matrix of the cavity at $B=0$, with the stub replaced by a regular
lead. The distribution of $U$ is the circular orthogonal ensemble (COE), which
is the ensemble of uniformly distributed, unitary and symmetric
matrices.\cite{Mehta} The distribution of $S(B)$ resulting from Eqs.\
(\ref{eq:VB}) and (\ref{eq:CUEB}) crosses over from the COE for $B=0$ to the
CUE for $B \to \infty$. One can show\cite{future} that it is equivalent to the
distribution of scattering matrices following from the Pandey-Mehta
Hamiltonian\cite{PandeyMehta} $H = H_0 + i B H_1$ [where $H_0$ ($H_1$) is a
real symmetric (antisymmetric) Gaussian distributed matrix].

It remains to relate the parameter $a$ to microscopic properties of the cavity.
We do this by computing the correlator $\Sigma_{mn} (\delta B) = \langle
S_{mn}^{\vphantom{*}}(B) S_{mn}^{*}(B + \delta B) \rangle$ from Eq.\
(\ref{eq:CUEB}). Using the diagrammatic method of Ref.\
\onlinecite{BrouwerBeenakker96} to perform the average over the COE, we find
(for $N+M \gg 1$)
\begin{eqnarray} \label{eq:Sx}
  \Sigma_{mn} &=& (N+M)^{-1} \left[1 + (\delta B/B_c)^2 \right]^{-1}, \ \ n
\neq m,
\end{eqnarray}
with $B_c \equiv  a^{-1} \sqrt{N+M}$. This correlator of scattering matrix
elements has also been computed by other
methods.\cite{JBS90,Pluhar,Efetov,Frahm} Comparing results we can identify
\begin{equation}
  a^2 = c e^2 v_F L^2 \min(\ell,L)/\hbar \delta,
\end{equation}
with $c$ a numerical coefficient of order unity depending on the shape of the
cavity (linear dimension $L$, mean free path $\ell$, Fermi velocity $v_F$,
level spacing $\delta$). For example, for a disordered disk or sphere (radius
$L \gg \ell$) the coefficient $c = \pi/8$ for the disk and $\pi/15 $ for the
sphere.

We now proceed with the calculation of the correlator of the conductance. We
consider broken time-reversal symmetry ($B \gg B_c$) and assume that $N$ and
$M$ are both $\gg 1$. Using the method of Ref.\ \onlinecite{BrouwerBeenakker96}
for the average over $U$, we obtain the average conductance $\langle G \rangle
= 2NM/(N+2M)$ and the correlator
\begin{eqnarray}\label{eq:NSresult}
 C(\delta B, \delta \phi) =
  { 16 K N^2 M^2 (N+M)^2} { (N+2M)^{-4}}  \nonumber \\  \mbox{} \times
  { (N+M)^2 + (N^2 + M^2 K) \cos^2(\delta \phi/2) \over
    \left[(N+M)^2 - M^2 K \cos^2(\delta \phi/2) \right]^2},
\end{eqnarray}
where we have abbreviated $K = \left[ 1 + (\delta B/B_c)^2 \right]^{-2}$.
Eq.\ (\ref{eq:NSresult}) simplifies considerably in the two limiting regimes $M
\ll N$ and $M \gg N$. For $M \ll N$ we find
\begin{mathletters} \label{eq:NS1}
\begin{eqnarray}
  C_0(\delta B) &=& 24 (M/N)^2 K, \\
  C_{\phi}(\delta B, \delta \phi) &=& 8 \left({M/N} \right)^2 K \cos \delta
\phi,
\end{eqnarray}
\end{mathletters}%
whereas for $M \gg N$ we have (for $| \delta \phi| < \pi$)
\begin{mathletters} \label{eq:NS2}
\begin{eqnarray}
  C_0(\delta B) &=& {\sqrt{N \over 8M}}  \left[1 + {M \over N} \left({\delta B
\over B_c}\right)^2 \right]^{-3/2}, \\
  C_{\phi}(\delta B, \delta \phi) &=& {1 \over 2} \left[1 + {M \over N}
\left({\delta B \over B_c}\right)^2 + {M \over 8N} {\delta \phi}^2\right]^{-2}.
\label{eq:NS2a}
\end{eqnarray}
\end{mathletters}%
The two regimes differ markedly in several respects:

(1) The $2\pi$-periodic conductance fluctuations are harmonic if $M \ll N$ and
highly anharmonic if $M \gg N$. A small increment $\delta \phi \simeq
\sqrt{N/M} \ll 2 \pi$ of the phase difference between the superconducting
contacts is sufficient to decorrelate the conductance if $M \gg N$.

(2) The variance of the conductance $\mbox{var}\, G = C_0(0) + C_{\phi}(0,0)$
has the universal magnitude $1/2$ if $M \gg N$, while it is reduced by a factor
$(8M/N)^2$ if $M \ll N$.

(3) The variance $\mbox{var}\, G_{\phi} = C_{\phi}(0,0)$ of the
$\phi$-dependent conductance is {\em larger} than the variance $\mbox{var}\,
G_0 = C_0(0)$ of the background conductance if $M \gg N$ (by a factor
$\sqrt{M/8N}$), while it is {\em smaller} if $M \ll N$ (by a factor $1/3$).

(4) The correlators $C_{\phi}(\delta B,0)$ and $C_0(\delta B)$ both decay as a
squared Lorentzian in $\delta B/B_c$ if $M \ll N$. If $M \gg N$, on the
contrary, $C_{\phi}(\delta B,0)$ decays as a squared Lorentzian, while
$C_0(\delta B)$ decays as a Lorentzian to the power $3/2$.

\begin{figure}[h]
\vspace{-0.5cm}

\hspace{0.05\hsize}
\epsfxsize=0.85\hsize
\epsffile{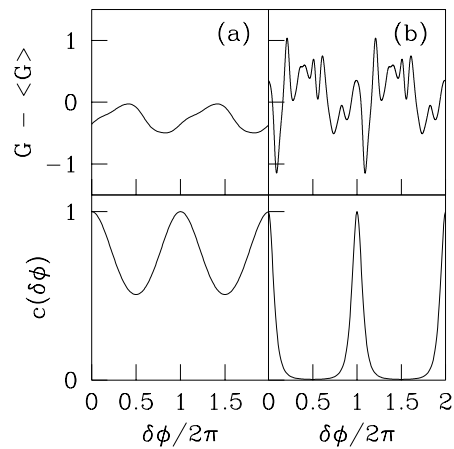}

\caption{\label{fig:3} Top panels: conductance minus the ensemble average (in
units of $2e^2/h$) as a function of the phase difference between the
superconductors. Bottom panels: normalized correlator $c(\delta \phi) =
C(0,\delta \phi)/C(0,0)$, computed from Eq.\ (\protect\ref{eq:NSresult}). (a)
is for $N=120$, $M=60$; (b) is for $N=10$, $M = 160$.}
\end{figure}

The difference between the two limiting regimes is illustrated in Fig.\
\ref{fig:3}. The ``sample-specific'' curves in the upper panels were computed
from Eq.\ (\ref{eq:NSconductance}) for a matrix $S$ which was randomly drawn
from the CUE. The correlators in the lower panels were computed from Eq.\
(\ref{eq:NSresult}). The qualitative difference between $M \lesssim N$ (Fig.\
\ref{fig:3}a) and $M \gg N$ (Fig.\ \ref{fig:3}b) is clearly visible.

We now turn to the four-terminal configuration (Fig.\ \ref{fig:1}b). The two
point contacts to the superconductors have $M/2$ modes each, as before; The two
point contacts to the normal metals have $N/2$ modes each. The conductance is
given by the four-terminal generalization of Eq.\
(\ref{eq:NSconductance}),\cite{Lambert}
\begin{mathletters}
\begin{eqnarray}
  && G = R_{21}^{ee} + R_{21}^{he} + {2(R_{11}^{he} R_{22}^{he} - R_{12}^{he}
R_{21}^{he}) \over R_{11}^{he} + R_{22}^{he} + R_{12}^{he} + R_{21}^{he}}, \\
  && R_{ij}^{he} = \mbox{tr}\, s_{he}^{\vphantom{\dagger}} c_j s_{he}^{\dagger}
c_i, \ \ R_{ij}^{ee} = \mbox{tr}\, s_{ee}^{\vphantom{\dagger}} c_j
s_{ee}^{\dagger} c_i, \\
  && s_{ee} =  r' + t e^{-i \Phi} r^{*} \left(1 + e^{i \Phi} r e^{-i \Phi}
r^{*} \right)^{-1} e^{i \Phi} t'.
\end{eqnarray}
\end{mathletters}%
Here $(c_1)_{mn} = 1$ if $m=n\le N/2$ and $0$ otherwise, and $c_2 = 1 - c_1$.
The matrix $s_{he}$ was defined in Eq.\ (\ref{eq:NSconductance2}). Performing
the averages as before, we find $\langle G \rangle = N/4$ and
\begin{eqnarray} \label{eq:NNresult}
  C(\delta B, \delta \phi) &=& \case{1}{16}
  {N^2 K [(N+M)^2 + M^2 K \cos^2(\delta \phi/2)]} \nonumber \\ && \mbox{}
\times
  {[(N+M)^2 - M^2 K \cos^2 (\delta \phi/2)]^{-2}}\label{eq:cov2}.
\end{eqnarray}
In the regime $M \ll N$ this simplifies to
\begin{mathletters} \label{eq:NN1}
\begin{eqnarray}
  C_0(\delta B) &=& \case{1}{16} K \\
  C_{\phi}(\delta B, \delta \phi) &=& \case{3}{32} \left({M/N} \right)^2 K^2
\cos \delta \phi,
\end{eqnarray}
\end{mathletters}%
while in the regime $M \gg N$ we find again Eq.\ (\ref{eq:NS2}) (with an extra
factor of $1/16$ on the r.h.s.).

The four-terminal configuration with $M \gg N$ is similar to the system studied
by Altshuler and Spivak.\cite{AltshulerSpivak} One basic difference is that
they consider the high-temperature regime $k_B T \gg \hbar/\tau_{\rm dwell}$
(with $\tau_{\rm dwell}$ the mean dwell time of a quasiparticle in the
junction), while we assume $T=0$ (which in practice means $k_B T \ll
\hbar/\tau_{\rm dwell}$). Because of this difference in temperature regimes we
can not make a detailed comparison with the results of Ref.\
\onlinecite{AltshulerSpivak}.

The features of the regime $M \lesssim N$ in the three-terminal configuration
agree qualitatively with the experimental observations made by Den Hartog et
al.\cite{Hartog} In particular, they find a nearly sinusoidal $\phi$-dependence
of the conductance, with $C_{\phi}(B,0)$ being smaller than $C_0(B)$, while
having the same $B$-dependence. The magnitude of the fluctuations which they
observe is much smaller than what we find for a point-contact coupling with $M$
and $N$ of comparable magnitude. This brings us to the prediction, that the
insertion of a point contact in the vertical arm of the T-junction of Ref.\
\onlinecite{Hartog} (which is connected to a normal metal) would have the
effect of (1) increasing the magnitude of the magnetoconductance fluctuations
so that it would become of order $e^2/h$; (2) introducing higher harmonics in
the $\phi$-dependence of the conductance. This should be a feasible experiment
which would probe an interesting new regime.

In conclusion, we have calculated the correlation function of the conductance
of a chaotic cavity coupled via point contacts to two superconductors and one
or two normal metals, as a function of the magnetic field through the cavity
and the phase difference between the superconductors. If the superconducting
point contacts dominate the conductance, the phase-dependent conductance
fluctuations are harmonic, whereas they become highly anharmonic if the normal
point contact limits the conductance. The harmonic regime has been observed in
Ref.\ \onlinecite{Hartog}, and we have suggested a modification of the
experiment to probe the anharmonic regime as well. We introduced a novel
technique to compute the magnetoconductance fluctuations, consisting in the
replacement of the magnetic field by a time-reversal-symmetry breaking stub.
This extension of the circular ensemble is likely to be useful in other
applications of random-matrix theory to mesoscopic systems.

We have benefitted from discussions with the participants of the workshop on
``Quantum Chaos'' at the Institute for Theoretical Physics in Santa Barbara, in
particular with B. L. Altshuler. Discussions with B.\ J.\ van Wees are also
gratefully acknowledged. This work was supported in part by the Dutch Science
Foundation NWO/FOM and by the NSF under Grant no.\ PHY94--07194.

\end{document}